\documentclass[aps,prl,superscriptaddress,reprint,floats,citeautoscript,floatfix,nobibnotes,nofootinbib,longbibliography]{revtex4-2}

\usepackage[intlimits,tbtags]{amsmath}
\usepackage{times}
\usepackage{color}
\usepackage{amssymb}
\usepackage{graphicx}
\usepackage{float}
\usepackage{bm}
\usepackage{array}
\usepackage{dcolumn}
\usepackage{setspace}
\usepackage{fancyhdr}
\usepackage{float}
\usepackage{placeins}
\usepackage{booktabs}
\usepackage{textcase}
\usepackage{multirow}

\newcolumntype{C}[1]{>{\centering\arraybackslash}p{#1}} 
\usepackage{graphicx}    
\usepackage{booktabs}   
\usepackage{setspace} 

\newcolumntype{d}[1]{D{.}{.}{#1} }
\usepackage{xcolor}

\usepackage{hyperref}
\hypersetup{
    colorlinks=true,       
    linkcolor=blue,        
    citecolor=blue,        
    urlcolor=blue,         
    breaklinks=true,       
    bookmarks=true,        
    pdfborder={0 0 0},     
}

\begin{document}
\title{Record-Breaking Elemental Superconductivity in Tetralayer Kagome Borophene}

\author{Yingnan Liu}\thanks{These two authors contributed equally to this work.}
\affiliation{State Key Laboratory of Integrated Optoelectronics and Key Laboratory of UV-Emitting Materials and Technology of Ministry of Education, School of Physics, Northeast Normal University, Changchun 130024, China}

\author{Yan Liu}\thanks{These two authors contributed equally to this work.}
\affiliation{Laboratory of Quantum Functional Materials Design and Application, School of Physics and Electronic Engineering, Jiangsu Normal University, Xuzhou 221116, China}

\author{Renyu Duan}
\affiliation{Laboratory of Quantum Functional Materials Design and Application, School of Physics and Electronic Engineering, Jiangsu Normal University, Xuzhou 221116, China}

\author{Menghui Wang}
\affiliation{Institute of Atomic and Molecular Physics, Jilin University, Changchun 130023, China}

\author{Meiling Xu}\email{Contact author: xml@calypso.cn}
\affiliation{Laboratory of Quantum Functional Materials Design and Application, School of Physics and Electronic Engineering, Jiangsu Normal University, Xuzhou 221116, China}

\author{Hanyu Liu}\email{Contact author: hanyuliu@jlu.edu.cn}
\affiliation{Key Laboratory of Material Simulation Methods and Software of Ministry of Education and State Key Laboratory of High Pressure and Superhard Materials, College of Physics, Jilin University, Changchun 130012, China}

\author{Shoutao Zhang}\email{Contact author: zhangst966@nenu.edu.cn}
\affiliation{State Key Laboratory of Integrated Optoelectronics and Key Laboratory of UV-Emitting Materials and Technology of Ministry of Education, School of Physics, Northeast Normal University, Changchun 130024, China}

\begin{abstract}
Superconductivity above the liquid-nitrogen temperature remains rare in two-dimensional elemental crystals, where strong covalent bonding often yields high phonon frequencies but insufficient electron--phonon coupling. Here, using first-principles calculations and fully anisotropic Migdal--Eliashberg theory, we predict tetralayer kagome borophene (TKB) stabilized by ABAB covalent stacking, as a liquid-nitrogen-temperature elemental superconductor. With a predicted critical temperature of $\sim$102 K, TKB sets a record-high value among previously reported elemental superconductors. Unlike known high-$T_c$ boron-based superconductors dominated by in-plane $\sigma$-bonding states and high-frequency in-plane B-B stretching modes, TKB realizes an out-of-plane $s$-$p_z$-bonding-mediated pairing mechanism, in which interlayer $s$-$p_z$ bonding states at the Fermi level are strongly coupled to low-frequency out-of-plane vibrations of boron atoms. These results reveal a distinct out-of-plane pairing channel in multilayer borophene and establish covalent stacking engineering as a potential route for high-$T_c$ superconductivity in two-dimensional materials.

\end{abstract}

\maketitle

Achieving high-temperature superconductivity in elemental materials remains a central challenge in condensed matter physics.~\cite{kamerlingh1911resistance} High pressure can substantially enhance superconductivity in bulk elemental solids;~\cite{sakata2011superconducting,zhang2022record,ishizuka2000pressure,shimizu2002superconductivity,struzhkin1997superconductivity,wang2023superconductivity,ying2023record} for example, scandium reaches a critical temperature ($T_c$) of $\sim$ 36 K at 260 GPa,~\cite{wang2023superconductivity,ying2023record} However, such superconducting states rely on extreme compression, limiting their accessibility, tunability, and device integration. Two-dimensionality provides a different route to high-$T_c$ elemental superconductivity. Unlike compressed bulk phases, two-dimensional (2D) elemental crystals can expose their bonding networks directly to external control, allowing their electronic structure and electron--phonon coupling (EPC) to be tuned by layer stacking, strain, charge doping, substrates, and interfaces.~\cite{oh2021evidence,chou2023kondo,wu2018theory,han2025signatures,TianYan2026125205} Therefore, realizing liquid-nitrogen-temperature superconductivity in an elemental 2D crystal would not only extend the $T_c$ limit of elemental superconductors under ambient conditions, but also provide a highly tunable platform for integrating superconductivity with nanoscale devices. Yet superconductivity above the liquid-nitrogen temperature remains exceptionally rare in 2D elemental systems, where high phonon frequencies are often not accompanied by sufficiently strong EPC.

\begin{figure*}[!t]
    \centering
    \includegraphics[width=1.0\linewidth]{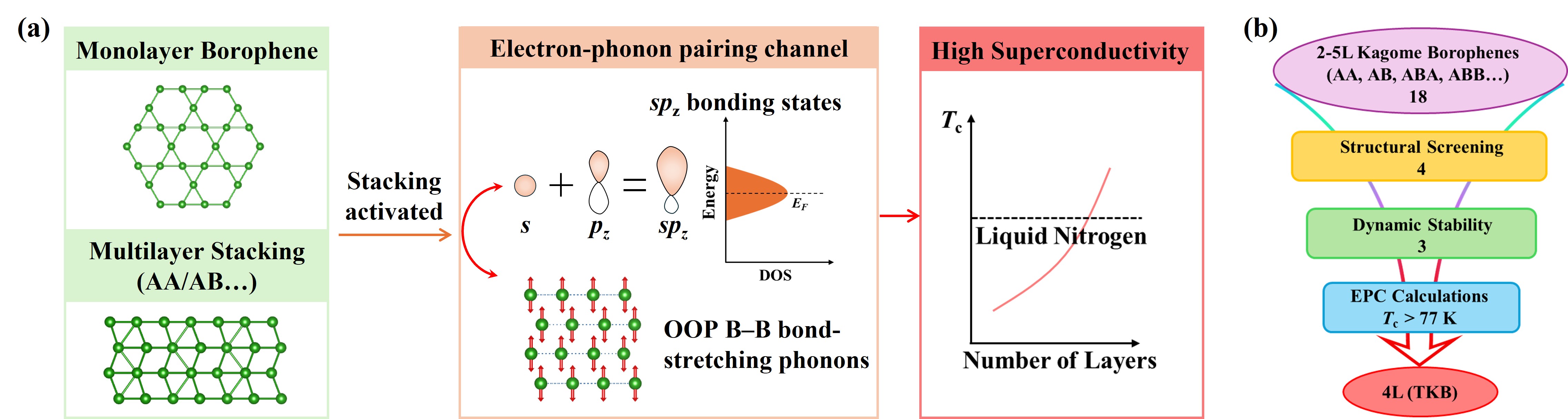}
    \caption{(a) Schematic illustration of the stacking-activated out-of-plane pairing mechanism. Multilayer stacking generates interlayer B-B covalent bonds and $s-p_z$ hybridized bonding states near the $E_F$, which couple strongly to out-of-plane (OOP) boron phonons and thereby enhance superconductivity beyond the liquid-nitrogen regime. (b) Screening workflow for multilayer kagome borophenes, including structural stability, dynamical stability, and EPC criteria.}
    \label{fig:1}
\end{figure*}

Borophene provides an attractive platform for addressing this challenge.~\cite{yan2021prediction,wang2024build,8l19-rdn2} The light mass, versatile covalent bonding, and rich structural polymorphism of boron endow high characteristic phonon frequencies and flexible electronic structures favorable for phonon-mediated pairing.~\cite{liu2025van} Several borophene phases have also been synthesized on metal substrates,~\cite{wu2019large,feng2016experimental,mannix2015synthesis} offering possible routes for experimental realization. Among them, kagome borophene is particularly appealing because its lattice geometry can host Dirac states, flat bands, and van Hove singularities near the Fermi level.~\cite{liu2025van,zhang2023realizing,qu2020boron,bo2020electron,zhang2025record,han2023two} However, in previously reported monolayer and few-layer borophenes, the electronic states relevant to superconductivity are mainly derived from in-plane B--B bonding networks, and the EPC remains insufficient to reach the liquid-nitrogen regime.\cite{yan2021prediction,gao2017prediction}

A key question is therefore whether stacking can do more than stabilize multilayer borophene or increase the density of states. In particular, can interlayer stacking activate a new pairing channel that is absent in monolayer borophene or weak in few-layer borophenes? This issue is especially important for boron-based superconductors, where high-$T_c$ pairing is commonly associated with in-plane $\sigma$-bonding states and in-plane B-B bond-stretching phonons.~\cite{an2001superconductivity,zhang2024conventional} An alternative route would be to create interlayer $s$-$p_z$ bonding states near the Fermi level ($E_F$) and couple them to low-frequency out-of-plane (OOP) boron vibrations, thereby combining strong covalent electronic states with softened phonon modes in a single elemental 2D system.

In this Letter, we predict tetralayer kagome borophene, stabilized by an ABAB covalent stacking configuration, as an elemental superconductor exhibiting a record-high $T_c$. The central idea, schematically illustrated in Figure~1(a), is to use stacking to activate an OOP electron-phonon pairing channel that is absent in monolayer borophene. ABAB stacking forms interlayer B-B covalent bonds and generates $s$-$p_z$ hybridized bonding states near the $E_F$, which are strongly coupled by low-frequency OOP boron phonons. Fully anisotropic Migdal--Eliashberg calculations yield a predicted $T_c$ of 102 K, setting a record-high value among previously reported intrinsic 2D elemental superconductors. These results reveal a distinct OOP $s$-$p_z$-bonding-mediated pairing mechanism and establish stacking engineering as a route to high-$T_c$ superconductivity in elemental 2D materials.


We first outline the design principle and structural screening strategy in Figure~1. As shown in Figure~1(a), multilayer stacking is expected to transform borophene from a predominantly in-plane bonded monolayer into a covalently stacked structure with interlayer B-B bonds. This stacking-activated bonding network gives rise to $s$-$p_z$-hybridized electronic states near the $E_F$, which can couple strongly to OOP B-B phonon modes, thereby providing an additional electron-phonon pairing channel. Guided by this principle, we screened 18 multilayer kagome borophene configurations with two to five layers and different stacking sequences [Figure~1(b); Figures  S1 and S2 within the Supporting Information]. This screening identifies the ABAB-stacked TKB as the optimal candidate for realizing stacking-activated electron-phonon pairing, with an estimated $T_c$ above the liquid-nitrogen temperature.

The optimized TKB structure [Figures ~2(a) and 2(b)] belongs to the trigonal space group $\textit{P}\overline{3}\textit{m}1$ and contains two inequivalent boron sites, B$_{\alpha}$ and B$_{\beta}$, which occupy the 6\textit{i} (0.827, 0.173, 0.479) and 6\textit{i} (0.164, 0.836, 0.436) Wyckoff sites, respectively. The interlayer B--B distances are 1.82 and 1.83~\AA{}, and are comparable to the intralayer B--B distances of 1.74--1.89~\AA{}. Electron localization function maps, electron localizability indicator, and localized orbital locator analyses confirm the strong in-plane and interlayer B-B covalent bonding ({Figures  S4-S6).

The energetic and mechanical stability of TKB is supported by cohesive-energy and elastic-constant calculations. Using isolated boron atoms as the reference, TKB has a cohesive energy of -6.52 eV/atom, lower than those of several experimentally synthesized borophene polymorphs, including $\delta_6$, $\alpha$, $\beta_{12}$, and $\chi_3$ borophene (-6.15 to -6.25 eV/atom).~\cite{mannix2015synthesis,zhao2016phonon,wu2019large,feng2016experimental} This indicates that TKB is energetically competitive with known borophene phases. In addition, the calculated elastic constants, $C_{11}=225.92$ N/m and $C_{12}=171.48$ N/m, satisfy the 2D Born stability criteria $C_{11}>0$ and $C_{11}>|C_{12}|$. The angle-dependent Young’s modulus and Poisson’s ratio reveal a nearly isotropic in-plane elastic response, further characterizing the mechanical robustness of TKB (Figure~S7 and Table~S2).

Four bands cross the Fermi level, confirm the intrinsic metallic character of TKB and give rise to multiple Fermi-surface sheets [Figure~2(c) and Figure~S8]. This metallicity is further supported by the atom- and orbital-resolved PDOS of B$_\alpha$ and B$_\beta$ atoms [Figure~2(e)]. In contrast to previously reported borophenes whose low-energy states are dominated mainly by in-plane $p_{x,y}$-derived orbitals,~\cite{yan2021prediction,zhang2025anisotropic,singh2001role,wang2024build} TKB exhibits substantial B $2s$ and $2p_z$ contributions near $E_F$, with the B $2p_z$ component extending prominently from about $-1.4$ eV to $E_F$. The corresponding partial charge density within this energy window is distributed over both intra- and interlayer B--B regions (Figure~S12), consistent with the mixed in-plane and OOP orbital characters revealed by the projected bands and PDOS. The orbital-resolved Fermi surfaces (FS) further reveal a clear orbital-selective topology [Figure~2(g)]. A flower-like hole pocket appears around the $\Gamma$ point and is mainly composed of B $s$ and $p_z$ orbitals, whereas the electron pockets near the K points and Brillouin-zone boundary are dominated by in-plane B $p_x$ and $p_y$ orbitals. Such momentum-dependent orbital differentiation provides an electronic basis for anisotropic EPC in TKB.

The bonding nature of the OOP $s$-$p_z$ channel is confirmed by the projected crystal orbital Hamilton population analysis [Figure~2(f)]. Compared with the B $2s$--$2p_x$ and $2s$--$2p_y$ interactions, the B $2s$--$2p_z$ component exhibits a pronounced bonding contribution near $E_F$, demonstrating the formation of interlayer $s$-$p_z$ bonding states in TKB.

In addition to this stacking-induced $s$-$p_z$ bonding channel, the kagome lattice produces characteristic electronic features near $E_F$. As shown in Figure~2(d), a van Hove singularity appears near the M point, accompanied by Dirac-like band crossings along the M--$\Gamma$ direction. These kagome-derived features enrich the anisotropic electronic landscape of TKB. A detailed analysis of the topological band features, including the Wannier-interpolated band structure, spin--orbit-coupling-included band structure, and edge spectral function, is provided in the SM (Figures  ~S10 and S11).

\begin{figure}
    \centering
    \includegraphics[width=1.0\linewidth]{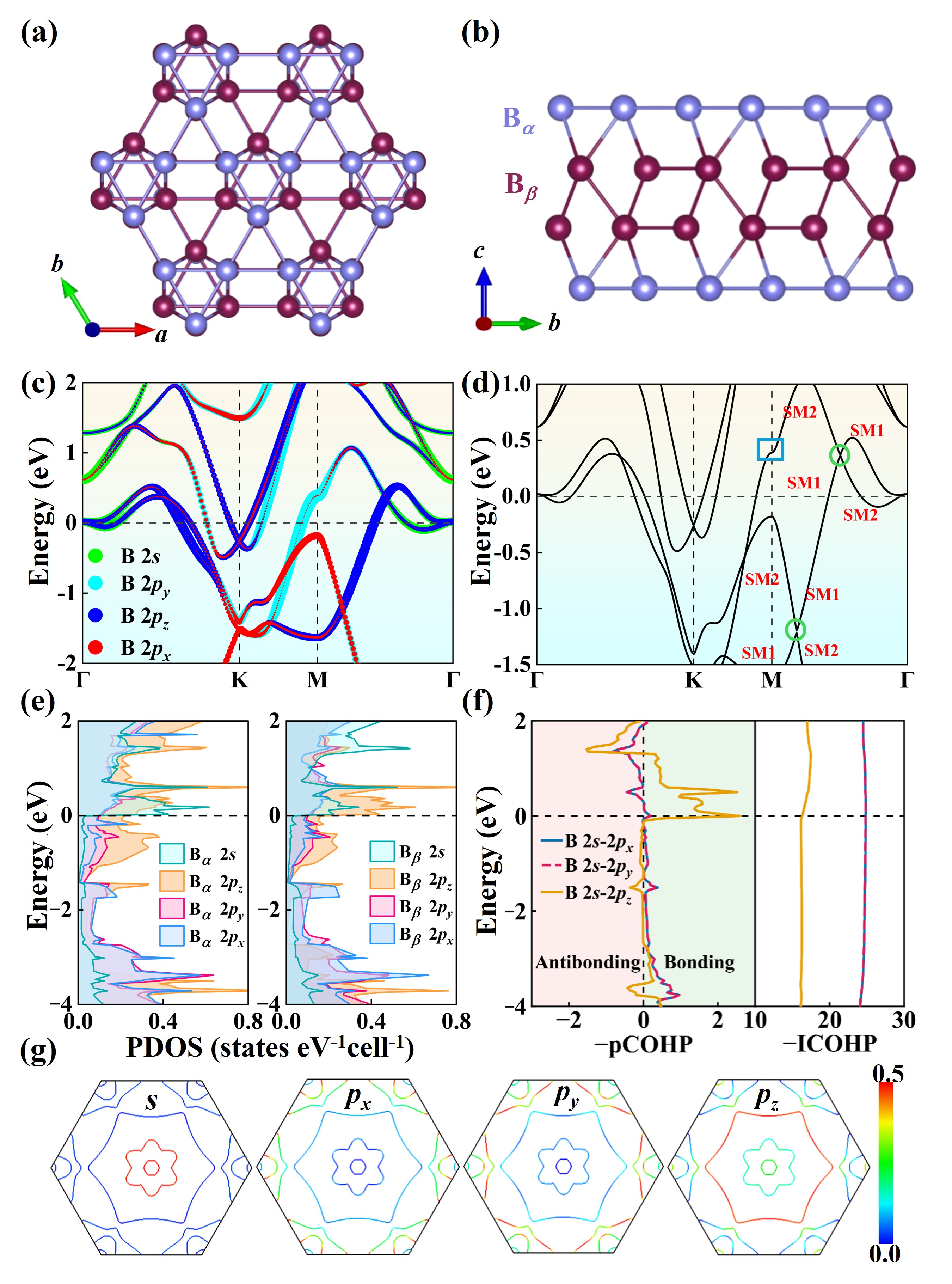}
    \caption{(a,b) Top and side views of the tetralayer kagome borophene structure, where the two colors denote inequivalent boron sites. (c) Orbital-resolved band structure, with the horizontal dashed line marking the Fermi level. (d) Enlarged view of the band crossing, with the corresponding irreducible representations labeled. The green circle marks the Dirac point (DP), while the blue rectangle highlights the Van Hove singularity. (e) Atom-resolved projected density of states (PDOS) for B$_\alpha$ and B$_\beta$ atoms. (f) Projected crystal orbital Hamilton populations (pCOHP) of 2$s$-$2p_x$, 2$s$-$2p_y$, and 2$s$-$2p_z$ pairs, together with the corresponding integrated COHP (ICOHP) in TKB. (g) Orbital-resolved FSs showing the contributions from $s$, $p_x$, $p_y$, and $p_z$ orbitals.}
    \label{fig:2}
\end{figure}

\begin{figure*}
    \centering
    \includegraphics[width=0.9\linewidth]{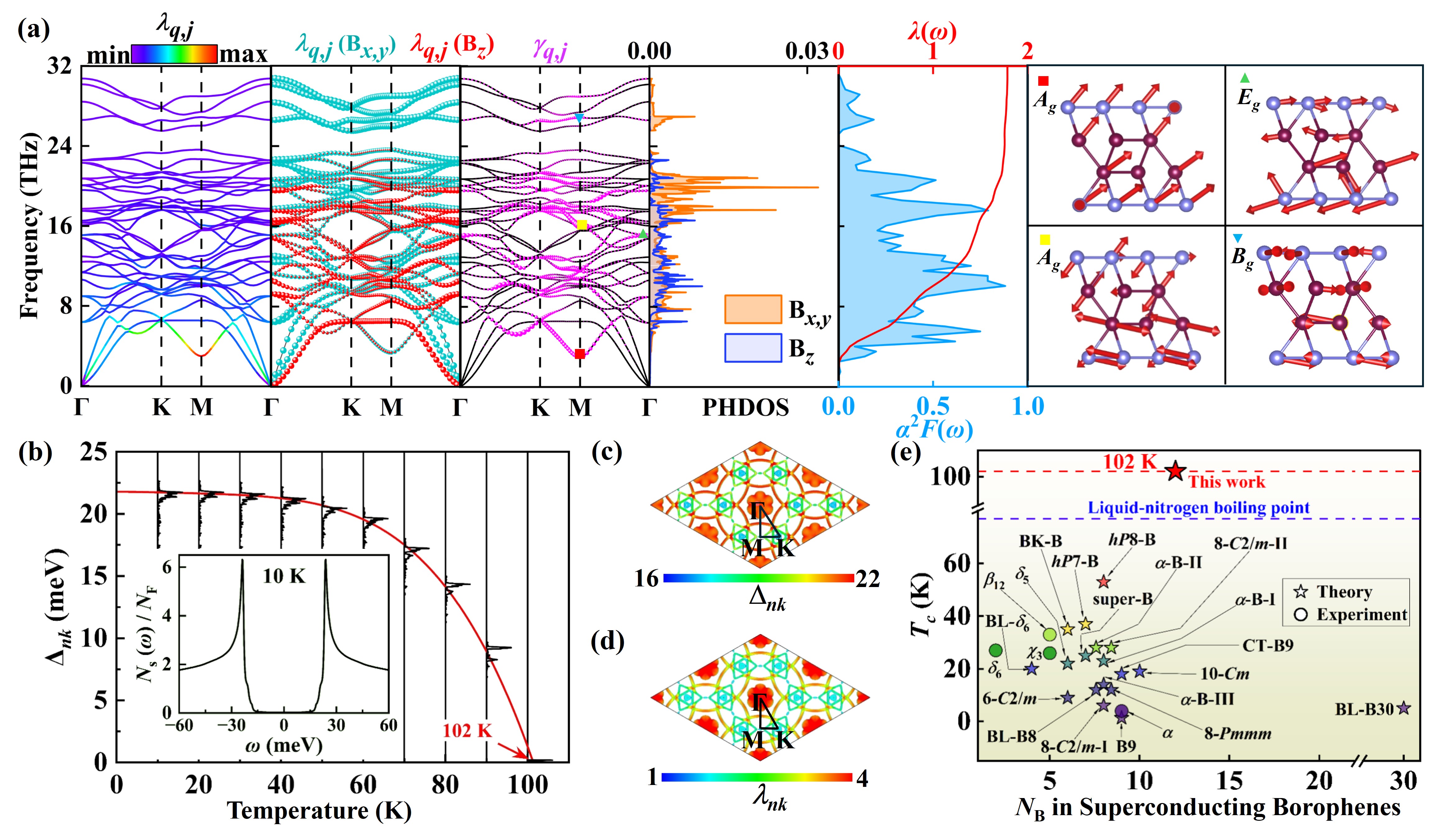}
    \caption{(a) Phonon dispersion weighted by the mode-resolved EPC strength $\lambda_{\mathbf{q}j}$, phonon branches associated with in-plane $B_{x,y}$ and out-of-plane $B_z$ vibrations, phonon dispersion weighted by the phonon linewidth $\gamma_{\mathbf{q}j}$, and frequency-dependent phonon density of states (PHDOS), Eliashberg spectral function $\alpha^2F(\omega)$, and accumulated EPC strength $\lambda(\omega)$. Four representative phonon modes with pronounced EPC contributions are also illustrated. (b) Temperature dependence of the superconducting gap $\Delta_{n\mathbf{k}}$. The inset shows the superconducting density of states (SDOS) at $T=10$ K. (c) Momentum-resolved superconducting gap $\Delta_{n\mathbf{k}}$ projected onto the FS at $T=10$ K. (d) Momentum-dependent EPC strength $\lambda_{n\mathbf{k}}$ projected onto the FS. (e) Comparison of $T_c$ values between TKB and previously reported borophene superconductors. The asterisks and circles denote theoretically predicted and experimentally synthesized borophenes, respectively. $N_{\rm B}$ denotes the number of boron atoms in the unit cell.}
\label{fig:3}
\end{figure*}

Having established the stacking-induced $s$-$p_z$ bonding states near $E_F$, we next examine how these electronic states couple to lattice vibrations. Using the Allen--Dynes modified McMillan equation~\cite{allen1975transition} with a Coulomb pseudopotential of $\mu^{*}=0.10$, we obtain an isotropic estimate of $T_c=65.7$ K and a total EPC strength of $\lambda=1.78$, indicating that TKB lies in the strong-coupling regime. To identify the microscopic origin of this large EPC, we calculated the phonon dispersion weighted by the mode-resolved EPC strength $\lambda_{\mathbf{qj}}$ and phonon linewidth $\gamma_{\mathbf{q}j}$, together with the phonon density of states, Eliashberg spectral function $\alpha^2F(\omega)$, and the accumulated EPC strength $\lambda(\omega)$ [Figure~3(a)].

The phonon spectrum shows pronounced EPC hot spots in the low-frequency region, especially near the M point and around 3 THz. The accumulated $\lambda(\omega)$ reveals that phonon modes below 17 THz contribute about 85\% of the total EPC, whereas the intermediate-frequency modes between 17 and 24 THz and the high-frequency modes above 24 THz contribute only 13\% and 2\%, respectively. Vibration-resolved analysis shows that both in-plane and OOP boron vibrations participate in the low-frequency EPC. In particular, the modes with large $\gamma_{\mathbf{qj}}$, including the $A_g$ mode near 3 THz at M and the $E_g$/$A_g$ modes around 15--16 THz, are dominated by out-of-plane $B_z$ vibrations of interlayer-bonding boron atoms. These modes strongly modulate the interlayer B--B bonding network. By contrast, the high-frequency $B_g$ mode around 27 THz at M is mainly associated with in-plane $B_{x,y}$ vibrations and contributes only weakly to the total EPC. These results demonstrate that the strong EPC in TKB is primarily driven by low-frequency phonons that couple efficiently to the stacking-induced $s$-$p_z$ bonding states at $E_F$, rather than by conventional high-frequency in-plane B--B stretching modes. 

We further solved the fully anisotropic Migdal--Eliashberg equations~\cite{margine2013anisotropic} to obtain a more reliable $T_c$. As shown in Figure~3(b), the temperature-dependent superconducting gap closes at $T_c=102$ K. The superconducting density of states at $T=10$ K exhibits well-defined coherence peaks, indicating an essentially single-gap superconducting state. We then examine the momentum dependence of the superconducting pairing. The momentum-resolved superconducting gap $\Delta_{n\mathbf{k}}$ and the momentum-dependent EPC strength $\lambda_{n\mathbf{k}}$ projected onto the FS are shown in Figures ~3(c) and 3(d), respectively. Both quantities exhibit pronounced anisotropy over the FS. Notably, the regions with large $\Delta_{n\mathbf{k}}$ and enhanced $\lambda_{n\mathbf{k}}$ coincide well with the Fermi-surface sheets carrying strong $s$-$p_z$ orbital weight [Figure~2(g)], directly linking the anisotropic superconducting gap to the stacking-induced OOP bonding states. This correlation demonstrates that the high-$T_c$ superconductivity in TKB is primarily driven by the coupling between $s$-$p_z$-derived Fermi-surface states and low-frequency OOP B--B phonon modes. 

As summarized in Figure~3(e), the predicted $T_c$ of TKB exceeds those of previously reported borophene superconductors,~\cite{liu2025van,yan2021prediction,zhao2016superconductivity,zhong2024superhard,wang2024build,zhang2025anisotropic,han2023two,li2018strain,zhao2018multigap} establishing ABAB-stacked TKB as a record-high-$T_c$ intrinsic 2D elemental superconductor. A broader comparison with representative elemental and 2D elemental superconductors is provided in Table~S3, further highlighting the exceptional superconductivity of TKB.~\cite{kamerlingh1911resistance,ying2023record,struzhkin1997superconductivity,ishizuka2000pressure,shimizu2002superconductivity,debessai2008comparison,zhang2022record,https://doi.org/10.1103/PhysRev.149.231,zhang2010superconductivity,cao2018unconventional,chen2019signatures,zhang2010superconductivity,yang2026metallenes} 

To further clarify why the tetralayer structure is optimal, we examined the electronic structures and EPC properties of bilayer and trilayer kagome borophenes (Figures ~S9 and S16). Their predicted $T_c$ values are only 0.7 and 8.1 K, respectively, much lower than that of TKB (Table~S4). Their suppressed superconductivity arises from a reduced density of states at $E_F$ and the absence of pronounced low-frequency phonon softening near the M point (Figure~S17), which together weaken the EPC associated with OOP B--B vibrations. By contrast, the five-layer kagome borophene becomes dynamically unstable, as indicated by imaginary phonon modes (Figures ~S2 and S3). Therefore, the high $T_c$ of TKB is not a simple layer-number effect; rather, the ABAB tetralayer configuration represents an optimal balance between structural stability and stacking-enhanced EPC. It uniquely combines sizable $s$-$p_z$ bonding states at $E_F$ with softened OOP B--B phonon modes near the M point, giving rise to the strong EPC responsible for the 102 K superconductivity. 

Since strain and carrier doping are commonly used to tune superconductivity in two-dimensional materials,~\cite{li2025emerging,wei2025electride} we further examined their effects on TKB. Phonon spectra calculations show that TKB becomes dynamically unstable under 1\% biaxial tensile strain, while it remains stable under 1\% biaxial compressive strain but becomes unstable when the compressive strain increases to 2\% (Figure~S18). These results indicate that TKB has a narrow strain-stability window. Moreover, under 1\% compressive strain, the $T_c$ is reduced to 59.6 K, as estimated using the Allen--Dynes modified McMillan equation (Figure~S18 and and Table~S5). Electron-doped TKB remains dynamically stable up to 0.10~e/cell; however, its $T_c$ decreases to 63.9 K (Figure~S18 and Table~S6). These results indicate that pristine TKB already lies close to the optimal electronic and structural configuration for high-$T_c$ superconductivity.

Inspired by the successful epitaxial growth of borophenes on suitable metal substrates, such as Ag(111) and Cu(111),~\cite{feng2016experimental,mannix2015synthesis,chen2022synthesis,liu2022borophene,kaneti2021borophene} we further assess the possible experimental realization of TKB on a substrate. As shown in Figure~S18, a $2\times2\times1$ TKB supercell can be matched with a $3\times3\times1$ Cu(111) slab with a lattice mismatch of $\sim3\%$, suggesting acceptable epitaxial compatibility. The calculated interfacial adhesion energy is $25~\mathrm{meV~\AA^{-2}}$, comparable to those reported for the bilayer borophenes $\beta_{12}$-B, $\chi_3$-B, and $\nu_{1/12}$-B synthesized on Ag(111). ~\cite{feng2016experimental,xu2022quasi} Electronic property analyses demonstrated that the B-derived density of states near $E_F$ is reduced after adsorption on Cu(111), likely due to substrate-induced tensile strain and interfacial electronic redistribution, while additional Cu-derived states appear around $E_F$ (Figures ~S20 and S21). These substrate effects may weaken the intrinsic EPC of TKB, as commonly encountered in substrate-supported borophenes.~\cite{gao2017prediction} Therefore, Cu(111) can serve as a viable template for epitaxial growth, whereas exfoliation or transfer from the substrate would be desirable for probing the intrinsic high-$T_c$ superconductivity predicted for freestanding TKB.

In summary, we predict ABAB-stacked TKB as an intrinsic 2D elemental superconductor with a record-high $T_c$. Fully anisotropic Migdal--Eliashberg calculations yield $T_c=$ 102 K for TKB, exceeding those of previously reported elemental superconductors. The high-$T_c$ state is driven by a stacking-activated OOP pairing channel, in which interlayer $s$-$p_z$ bonding states at $E_F$ couple strongly to low-frequency OOP vibrational modes. This mechanism is distinct from conventional boron-based superconductors dominated by in-plane $\sigma$-bonding states and high-frequency in-plane B--B stretching modes. These findings identify covalent stacking as a design principle for realizing high-$T_c$ superconductivity beyond conventional in-plane bonding mechanisms in 2D materials.

\noindent\textbf{ASSOCIATED CONTENT}\\
\textbf{Supporting Information}\\
The Supporting Information is available free of charge at https://pubs.acs.org/doi/10.1021/XXX. \\
Computational details; structures and stability of multilayer kagome borophenes; electronic structure and bonding of TKB; electronic and superconducting properties of TrKB; phonon spectra of TKB under strain and doping; structure and properties of TKB/Cu(111) (PDF).\\

\noindent\textbf{AUTHOR INFORMATION}\\
\textbf{Corresponding Authors}\\
\textbf{Meiling Xu} - Laboratory of Quantum Functional Materials Design and Application, School of Physics and Electronic Engineering, Jiangsu Normal University, Xuzhou 221116, China; https://orcid.org/0000-0001-6592-8975; E-mail: xml@calypso.cn\\
\textbf{Hanyu Liu} - Key Laboratory of Material Simulation Methods and Software of Ministry of Education and State Key Laboratory of High Pressure and Superhard Materials, College of Physics, Jilin University, Changchun 130012, China; https://orcid.org/0000-0003-2394-5421; E-mail: hanyuliu@jlu.edu.cn\\
\textbf{Shoutao Zhang} - Key Laboratory of UV-Emitting Materials and Technology of Ministry of Education, School of Physics, Northeast Normal University, Changchun 130024, China; https://orcid.org/0000-0002-0971-8831; E-mail: zhangst966@nenu.edu.cn\\

\noindent\textbf{Authors}\\
\textbf{Yingnan Liu} - Key Laboratory of UV-Emitting Materials and Technology of Ministry of Education, School of Physics, Northeast Normal University, Changchun 130024, China; https://orcid.org/0009-0007-6195-8158\\
\textbf{Yan Liu} - Laboratory of Quantum Functional Materials Design and Application, School of Physics and Electronic Engineering, Jiangsu Normal University, Xuzhou 221116, China\\
\textbf{Renyu Duan} - Laboratory of Quantum Functional Materials Design and Application, School of Physics and Electronic Engineering, Jiangsu Normal University, Xuzhou 221116, China\\
\textbf{Menghui Wang} - Institute of Atomic and Molecular Physics, Jilin University, Changchun 130023, China\\

\noindent\textbf{Author Contributions}\\
Yingnan Liu and Yan Liu contributed equally to this work.\\

\noindent\textbf{Notes}\\
The authors declare no competing financial interest.\\

\noindent\textbf{ACKNOWLEDGMENTS}\\
This work was supported by the National Natural Science Foundation of China (Grant No. 11704062, No. 12474012, and No. 12574012), the Science and Technology Development Plan Project of Jilin Province, China (Grant No. 20260102248JC), and the “111 Center” (No. B25030).\\

\bibliography{References}

\end{document}